# Experimental demonstration of a hyper-entangled ten-qubit Schrödinger cat state

Wei-Bo Gao[1], Chao-Yang Lu[1], Xing-Can Yao[1], Ping Xu[1], Otfried Gühne[2,3], Alexander Goebel[4], Yu-Ao Chen[1,4], Cheng-Zhi Peng[1], Zeng-Bing Chen[1], Jian-Wei Pan[1,4,*]

(1) *Hefei National Laboratory for Physical Sciences at Microscale and Department of Modern Physics, University of Science and Technology of China, Hefei, Anhui 230026, P.R. China;* (2) *Institut für Quantenoptik und Quanteninformation, Österreichische Akademie der Wissenschaften, Technikerstraße 21A, A-6020 Innsbruck, Austria;* (3) *Institut für Theoretische Physik, Universität Innsbruck, Technikerstraße 25, A–6020 Innsbruck, Austria.* (4) *Physikalisches Institut, Universität Heidelberg, Philosophenweg 12, D-69120 Heidelberg, Germany.*
\* email: jian-wei.pan(at)physi.uni-heidelberg.de

**Abstract:**

Coherent manipulation of an increasing number of qubits for the generation of entangled states has been an important goal and benchmark in the emerging field of quantum information science. The multiparticle entangled states serve as physical resources for measurement-based quantum computing *(1)* and high-precision quantum metrology *(2,3)*. However, their experimental preparation has proved extremely challenging. To date, entangled states up to six *(4)*, eight *(5)* atoms, or six photonic qubits *(6)* have been demonstrated. Here, by exploiting both the photons' polarization and momentum degrees of freedom, we report the creation of hyper-entangled six-, eight-, and ten-qubit Schrödinger cat states. We characterize the cat states by evaluating their fidelities and detecting the presence of genuine multi-partite entanglement. Small modifications of the experimental setup will allow the generation of various graph states up to ten qubits. Our method provides a shortcut to expand the effective Hilbert space, opening up interesting applications such as quantum-enhanced super-resolving phase measurement, graph-state generation for anyonic simulation *(7)* and topological error correction *(8,9)*, and novel tests of nonlocality with hyper-entanglement *(10,11)* .

Linear optical control of single photonic qubits has been an appealing approach to implementations of quantum computing *(12,13)*. Experiments in recent years have demonstrated the photons' extremely long decoherence time *(14)*, fast clock speed *(15)*, a series of controlled quantum logic gates *(16-19)* and algorithms *(15,20-22)*, and the generation of various multiqubit entangled states *(6,23)*. A significant challenge lies on, however, the experimentally accessible source of multiple photonic qubits. This is because, on the one hand, the probabilistic nature of spontaneous parametric down conversion *(24)* puts a bottleneck on the brightness and fidelity of multiphoton states; manipulating more than six or seven photons without quantum storage seems an insurmountable task. On the other hand, triggered single-photon sources *(25)* from independent quantum dots or other emitters still suffer from spectral distinguishablity at present which prevents them from the ready scalability.

There is, however, a shortcut to experimentally control more effective qubits by exploiting the hyper-entanglement *(26)* – the simultaneous entanglement in multiple degrees of freedom which naturally exist for various physical systems. For instance, one can not only encode quantum information by the polarization of a single photon, but also by its spatial modes *(10)*, arrival time *(27)* or orbital angular momentum *(28)*. Recently, hyper-entangled photonic states *(29)* have been experimentally realized, and shown to offer significant advantages in quantum super-dense coding *(30,31)*, enhanced violation of local realism *(32,33)*, efficient construction of cluster states *(34,35)* and multi-qubit logic gates *(19)*.

Although the largest hyper-entangled state *(29)* realized so far has expanded the Hilbert space up to an impressive 144-dimensional, it is a product state of two-party entangled states and does not involve multipartite entanglement. Other schemes *(34,35)* for creating hyper-entanglement have been limited by the technical problem of photonic sub-wavelength phase stability and appear infeasible to generate larger states than the four-qubit ones. In this Letter we will describe our method which overcomes these limitations, and the experimental generation of hyper-entangled six-, eight-, and ten-qubit photonic Schrödinger cat states.

The Schrödinger cat states, also technically known as Greenberger-Horne-Zeilinger states *(36)*, involve an equal superposition of two maximally different quantum states. They are of particular interest in quantum mechanics and find wide applications in quantum information processing *(see a review e.g. ref. 37)*. Our experiment aims to create the cat state in the form:

$$|Cat\rangle^{2n} = (|H\rangle^{\otimes n}|H'\rangle^{\otimes n} + |V\rangle^{\otimes n}|V'\rangle^{\otimes n})/\sqrt{2} \qquad (1)$$

where $H$ and $V$ denote horizontal and vertical polarization, and $H'$ and $V'$ label two orthogonal spatial modes (or momentums) of the photons. The state (1) exhibits maximal entanglement between all photons' polarization and spatial qubits.

Our first experimental step is to generate polarization-entangled $n$-qubit cat states $|Cat\rangle_p^n = (|H\rangle^{\otimes n} + |V\rangle^{\otimes n})/\sqrt{2}$ by combining entangled photon pairs produced by spontaneous parametric down-conversion *(24)* and a pseudo-single photon source *(38)*. As shown in Fig.1**a**, a femtosecond infrared (IR) laser is attenuated to be a weak coherent photon source which has a very small probability ($p \sim 0.03$) of containing a single photon for each pulse, and prepared in the superposition state $|\psi\rangle = (|H\rangle + |V\rangle)/\sqrt{2}$ in path 1. Meanwhile, a pulsed ultraviolet (UV) laser, which is up-converted from the IR laser, passes through two $\beta$-barium borate (BBO) crystals generating two pairs of entangled photons in the state $|\phi\rangle = (|H\rangle|H\rangle + |V\rangle|V\rangle)/\sqrt{2}$ in path 2-3 and 4-5. The photo pairs have an average two-photon coincidence count rate of $2.4 \times 10^4 / s$ and a visibility of 0.92 in the $H/V$ basis and 0.90 in the $(H \pm V)/\sqrt{2}$ basis. Photons from path 3 and 4 are superposed on a polarizing beam splitter (PBS$_1$), and then are further combined with photon from path 1 on PBS$_2$ (see Fig.1**a**). Fine adjustments of the delay between the different paths are made to ensure that the photons arrive at the PBSs simultaneously. Further, the photons are spectrally filtered and detected by single-mode fiber-coupled single-photon detectors for good spatial and temporal overlap. Since the PBSs transmit $H$ and reflect $V$ polarization, it can be concluded that a coincidence detection of the five output photons implies that all the photons are either $H$ or $V$ polarized – two cases quantum mechanically indistinguishable – thus projecting them in the cat state $|Cat\rangle_p^5 = (|H\rangle^{\otimes 5} + |V\rangle^{\otimes 5})/\sqrt{2}$. It is easy to check that, in a similar way, if we only combine photon 1 and 4 (3 and 4) on the PBS$_2$ (PBS$_1$), entangled cat states between the three photons 1-4-5 (the four photons 2-3-4-5) can be created.

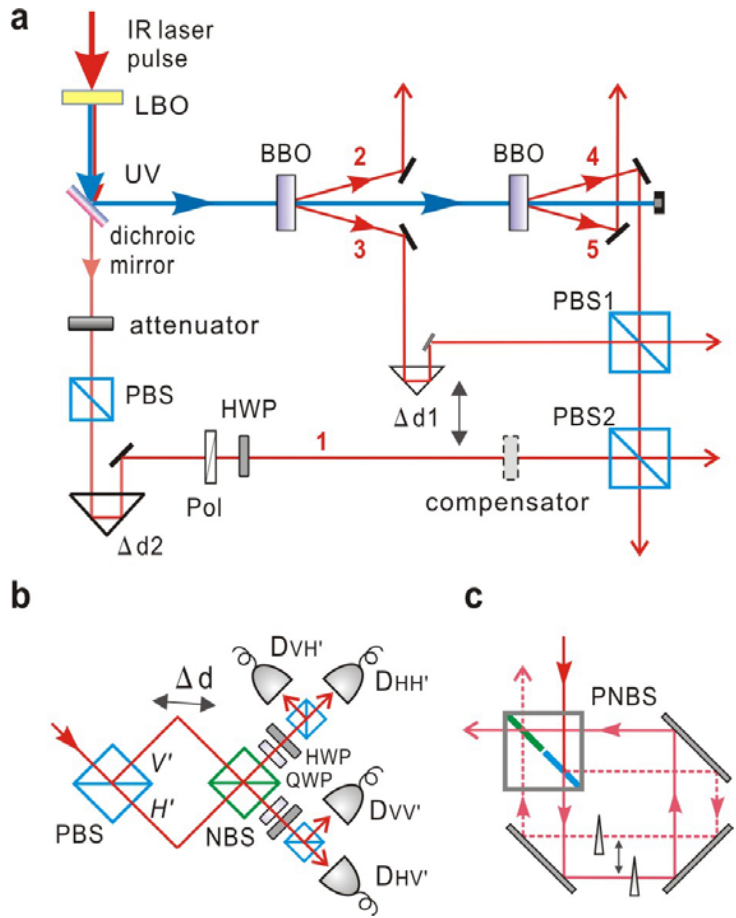

**Figure 1** Experimental setup for the generation of hyper-entangled six-, eight- and ten-qubit Schrodinger cat states. **a**. A mode-locked Ti:Sapphire laser outputs a pulsed infrared (IR) laser with a central wavelength of 788 nm, a pulse duration of 120 fs and a repetition rate of 76 MHz, which passes through a LBO ($LiB_3O_5$) crystal and converted to UV laser ($\lambda = 394nm$). Behind the LBO, five dichroic mirrors (only one shown) are used to separate the mixed UV and IR light components. The UV laser is focused on two BBO crystals to produce two pairs of entangled photons. The transmitted IR laser is attenuated to a weak coherent photon source. After the photons are overlapped on the two PBSs (see *Text*), they are guided out to the setup shown in **c**. Finally they are detected by fibre-coupled single-photon detectors and the coincidence events are registered by a programmable multi-channel coincidence unit. **b**. Conceptual interferometer for implementation and analysis of hyper-entanglement. A incoming photon is split into two possible spatial modes by the PBS with regard to its polarization: $H \rightarrow H'$, $V \rightarrow V'$, forming an EPR-like entangled state between the photons' spatial and polarization degrees of freedom. A Mach-Zehnder-type interferometer with a non-polarizing beam splitter (NBS) is used to coherently measure the spatial mode qubit, and subsequently at both of its output port conventional polarization analysis is performed. **c**. The experimentally stable interferometer with a Sagnac-like configuration. The specially designed beam splitter cube (PNBS) is half-PBS coated and half-NBS coated. High-precision small-angle prisms are inserted for fine adjustments of the relatively delay of the two different paths.

Next, we grow the polarization-encoded *n*-qubit states into double-sized 2*n*-qubit cat states by planting spatial modes on them. This again, exploits PBSs. Consider a polarized single photon qubit in the state $\alpha|H\rangle + \beta|V\rangle$ passes through a PBS (see

Fig.1**b**). The PBS separates the photon into two possible spatial modes $H'$ and $V'$ according to their polarization $H$ and $V$ respectively; indeed, this forms the basis of the PBS as an instrument for measuring polarization. The state of this single photon can now be written as $\alpha|H\rangle|H'\rangle + \beta|V\rangle|V'\rangle$, an entangled state between its polarization and spatial degree of freedom. It is straightforward to extend this method on the *n*-photon state $|Cat\rangle_p^n$; thereby the hyper-entangled 2*n*-qubit cat state (1) can be created. Besides the cat states, we note that this method can also be flexibly modified for the generation of other graph states *(39)* which are central resources in measurement-based quantum computing (see *Appendix Fig. A1*).

With multiple degrees of freedom carrying the quantum information in a single photon, measurements of the composite quantum states now become a bit trickier, as it is necessary to read out one degree of freedom without destroying another one. Illustrated also in Fig. 1**b** is the followed-up apparatus for simultaneously measuring both the polarization and spatial qubits on the basis of $|0\rangle/|1\rangle$ and $(|0\rangle \pm e^{i\theta}|1\rangle)$ (here we denote $|H\rangle$ and $|H'\rangle$ as logic $|0\rangle$, $|V\rangle$ and $|V'\rangle$ as $|1\rangle$). Specifically, the measurement of the spatial qubit employs an optical interferometer combing the two paths onto a non-polarizing beam splitter (NBS) with adjustable phase delay between these two paths. After this interferometer, the polarization information is then read out by placing a combination of a quarter-wave plate (QWP), a half-wave plate (HWP) and a PBS in front of the single-photon detectors. Experimentally, however, it is difficult to directly implement the interferometer in Fig. 1**b**, because it is sensitive to path length instability on the order of the photon's wavelength. To overcome this problem, we construct intrinsically stable Sagnac-like interferometers with beam-splitter cubes that are half PBS-coated and half NBS-coated (see Fig. 1**c**). The long-term stabilities and the high visibilities of the five interferometers constructed in our experiment are shown in *Appendix* Fig. A2.

As a step-by-step approach, we begin with the creation of the hyper-entangled six-qubit cat state $|Cat\rangle^6$ and eight-qubit cat state $|Cat\rangle^8$. To analyze the experimentally produced states, we first look at the measurement results in the $|0\rangle/|1\rangle$ basis as shown in Fig. 2**a** and Fig. 2**c** for $|Cat\rangle^6$ and $|Cat\rangle^8$ respectively. For ideal cat states, the desired combinations in this basis should in principle be $|H\rangle^{\otimes n}|H'\rangle^{\otimes n}$ and $|V\rangle^{\otimes n}|V'\rangle^{\otimes n}$ only. This is confirmed by the experiment data shown in Fig.2**a, c** showing obviously that these two terms dominate the overall coincidence events, with a signal-to-noise ratio (defined as the ratio of the average of the desired components to that of the other non-desired ones) of 85:1 to 1100:1 for the state $|Cat\rangle^6$ and $|Cat\rangle^8$ respectively. We note the undesired noise, noticeably located in the diagonal line of Fig. 2**a**, **c**, mainly arises from the double pair emission of entangled photons.

While the above data has determined the population in the $|0\rangle/|1\rangle$ basis of the cat states, it is not sufficient to reveal their coherence properties. Now we take measurements in the basis of $|R\rangle = (|0\rangle + e^{i\theta}|1\rangle)/\sqrt{2}$ and $|L\rangle = (|0\rangle - e^{i\theta}|1\rangle)/\sqrt{2}$. In this new basis, the cat state $|Cat\rangle^n$ can be written in the form of $(|R\rangle + |L\rangle)^{\otimes n} + e^{in\theta}(|R\rangle - |L\rangle)^{\otimes n}$ thus it becomes clear that the probability density of the components: $|R\rangle^{\otimes n}$, $|R\rangle^{\otimes(n-1)}|L\rangle$,

$\cdots|L\rangle^{\otimes n}$, hence the corresponding experimentally expected coincidence events, should obey the relation $\propto (1\pm\cos n\theta)$. From these measurements, one can determine the expectation values of spin observable: $\langle M_\theta^{\otimes n}\rangle = \cos n\theta$, where $M_\theta = \cos\theta\sigma_x + \sin\theta\sigma_y$, which oscillates $n$-times sinusoidally over a single cycle of $2\pi$. Indeed, this can only arise from coherent superposition between the $|0\rangle^{\otimes n}$ and $|1\rangle^{\otimes n}$ component of the cat state and serve as a characteristic signature of $n$-qubit coherence.

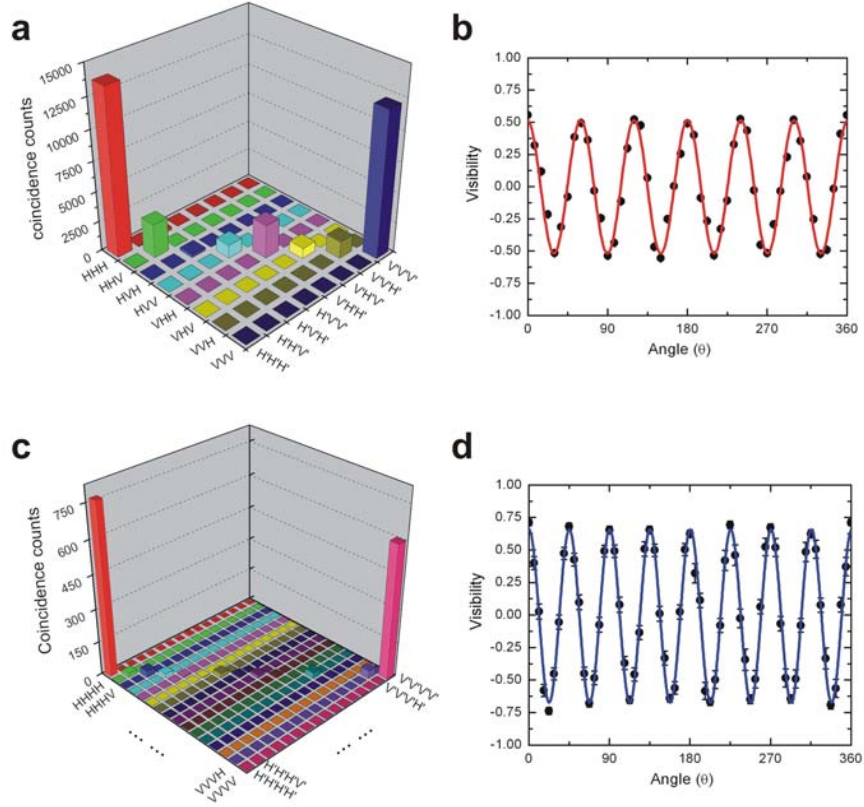

**Figure 2** Experimental results for determination of the fidelities of the six- and eight-qubit cat states. **a**, **c**. Coincidence counts obtain in the $|H\rangle/|V\rangle$ ($|H'\rangle/|V'\rangle$) basis, accumulated for 150s and 480s for the six- (**a**) and eight-qubit (**b**) state respectively. **b**, **d**. Measurement results obtained along the basis of $(|H\rangle + e^{i\theta}|V\rangle)/\sqrt{2}$ (($|H'\rangle + e^{i\theta}|V'\rangle)/\sqrt{2}$), showing the coherence of the cat states and phase super-resolution (see *Text*). The error bar stands for one standard deviation (in **b** it is too small to be seen) deduced from propagated Poissonian counting statistics of the raw detection events.

Figure 2**b**, **d** show the experimentally obtained expectation values $\langle M_\theta^{\otimes n}\rangle$ as a function of $\theta$ ($0 \le \theta \le 2\pi$) and the fitted sinusoidal fringes. The fringes clearly exhibit the $n\theta$ oscillation, with a visibility of $0.527\pm0.002$ and $0.67\pm0.1$ for the six- and eight-qubit cat state respectively, confirming the coherence between all effective $n$ qubits encoded with either polarization or spatial information. We note that the reduction of the visibilities is caused by, besides the above mentioned double pair photon emission, also the imperfections of photon overlapping at the PBSs and NBSs.

From the data shown in Fig. 2, we can further determine the fidelities of the cat states and detect the presence of genuine multipartite entanglement *(40,41)*. The fidelity – a measure of to what extent the desired state is created – is judged by the overlap of the experimentally produced state with the ideal one: $F(|\psi\rangle) = \langle\psi|\rho_{exp}|\psi\rangle$. For the cat state, $|\psi\rangle\langle\psi|$ can be decomposed as $(1/2)[(|0\rangle\langle 0|)^{\otimes n} + (|1\rangle\langle 1|)^{\otimes n} + (1/n)\sum_{k=1}^{n}(-1)^k M_{(k\pi/n)}^{\otimes n}]$, corresponding to measurements in the basis of $|0\rangle/|1\rangle$ and $(|0\rangle \pm e^{i(k\pi/n)}|1\rangle)$. Figure 2**a-d** shows the experimental results, from which the fidelities of the six- and eight-qubit cat state can be determined: $F(|Cat\rangle^6) = 0.6308 \pm 0.0015$, $F(|Cat\rangle^8) = 0.776 \pm 0.006$. The notion of genuine multipartite entanglement characterizes whether generation of the state requires interaction of all parties, distinguishing the experimentally produced state from any incompletely entangled state. For the cat states, it is sufficient to prove the presence of genuine multipartite entanglement *(40,41)* if their fidelities exceed the threshold of 0.5. Thus, with high statistical significance, the genuine *n*-qubit entanglement of the cat states created in our experiment is confirmed. We notice that the fidelity of the six-qubit cat state is considerably lower than that of the eight-qubit state, which is due to the fact that the generation of the former involves a faint coherent laser light which introduces more noise than the configuration of the latter. It is worth to mention here that an advantage the hyper-entanglement brings is that, our new six-qubit cat state not only has a higher fidelity than the previous six-photon cat state *(6)*, but also its count rate reaches $\sim 200/s$, an impressive $\sim 4$ orders of magnitude brighter than the six-photon coincidence.

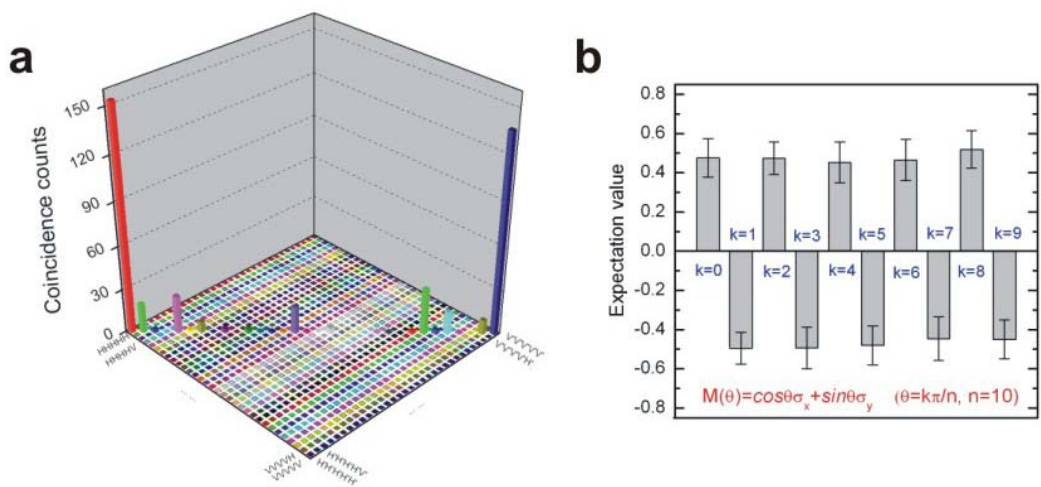

**Figure 3** Experimental results for the ten-qubit cat states. **a**. Coincidence counts measured in the $|H\rangle/|V\rangle$ ($|H'\rangle/|V'\rangle$) basis for 6 hours. **b**. The expectation values of $M_\theta^{\otimes 10}$, each derived from a complete set (each with 1024 different settings) of ten-qubit coincidence events in 1.5 hours in the basis of $|H\rangle \pm e^{ik\pi/10}|V\rangle$ ($|H'\rangle \pm e^{ik\pi/10}|V'\rangle$).

Now we proceed with the generation and analysis of the ten-qubit cat state $|Cat\rangle^{10}$ which employs the full setup depicted in Fig.1. Because of the probabilistic nature of spontaneous parametric down-conversion, the coincidence count rate of the ten-qubit state is as low as 0.021Hz, 1/160 (1/11000) of that of the eight-qubit (six-qubit) state. Measurements results along the $|0\rangle/|1\rangle$ basis are shown in Fig.3**a** with all $2^{10} = 1024$ possible combinations ($|H\rangle^{\otimes 5}|H'\rangle^{\otimes 5}$, $|H\rangle^{\otimes 5}|H'\rangle^{\otimes 4}|V'\rangle$, $\cdots|V\rangle^{\otimes 5}|V'\rangle^{\otimes 5}$) plotted, giving a signal to noise ration (defined the same as before) of 940:1. To determine the fidelity of the ten-qubit cat state, we further take measurements in the $(|0\rangle \pm e^{i\theta}|1\rangle)$ basis, where $\theta$ is chosen as: $\theta = k\pi/10$, $k = 0, 1, \ldots 9$. The measured expectation values of the observable $M_\theta^{\otimes 10}$ are listed in Fig.3**b** with an average absolute value of 0.475 – this can also be seen equivalently as the fringe visibility displayed in Fig. 2**b**, **d**. We can thus calculate the state fidelity: $F(|Cat\rangle^{10}) = 0.561 \pm 0.019$, which is above the threshold of 0.5 by more than 3 standard deviations, thus establishing the presence of genuine ten-qubit entanglement. These data is further analyzed using an optimized entanglement witness method (see *Appendix*) which, with even higher significance, confirms that the entanglement truly involves all ten qubits.

In summary, we have experimentally generated and analyzed the hyper-entangled six-, eight- and ten-qubit photonic Schrödinger cat state. These results represent the largest entangled state realized so far, expanding the effective Hilbert space up to 1024 dimensions. The cat states demonstrated here, together with other graph states technically feasible within our experimental method (see *Appendix Fig. S1*), create a versatile testing ground for study of multipartite entanglement and quantum information protocols. Indeed, by taking advantage of the hyper-entanglement, it is now possible to reach some experimental regimes that are hardly accessible before, for instance, demonstrations of the robustness of anyonic braiding *(7)*, topological or decoherence-free cluster-state encoding *(8,9,42)* that require manipulation of 7-10 qubits. It will be interesting in the future work to further exploit quantum particles' other degrees of freedom, such as arrival-time *(27)* or orbital angular momentum *(28)*, to create larger multi-dimensional entangled states for more efficient quantum information processing.

**Acknowledgments**: This work was supported by the National Natural Science Foundation of China, the Chinese Academy of Sciences and the National Fundamental Research Program (under Grant No: 2006CB921900). This work was also supported by the Alexander von Humboldt Foundation, the ERC, the FWF (START prize), and the EU (SCALA, OLAQUI, QICS).

# Appendix

**Figure A1**

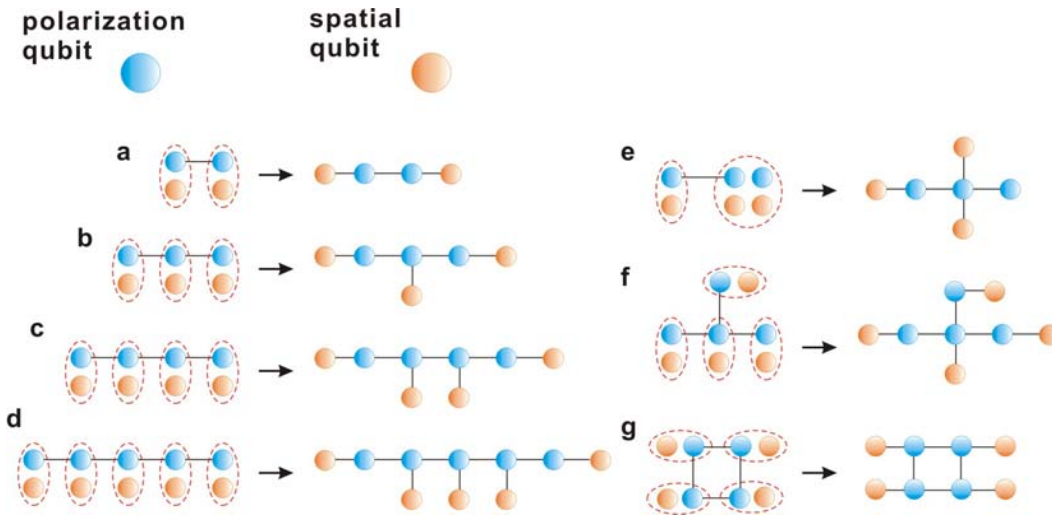

Figure A1: A few examples of how hyper-entangled photonic graph states can be generated. Adding a spatial qubit to a polarization qubit in the way described in our experiment can be viewed as redundantly encoding, doubling the effective qubits in the graph state (see *e.g.* **a-d**). Different types of hyper-entangled graph states may be created simply by local unitary transformations (see **b-e**, **c-g**), or by varying initial polarization-encoded graph states; for instance, staring from a four-photon linear cluster state (see **c**) or GHZ state (see **f**). Therefore, a large variety of graph states are available. The method of generating larger graph states from smaller ones has been proposed for polarization-encoded photonic states and referred to as "fusion of qubits" in *ref.* [1,2].

**Figure A2**

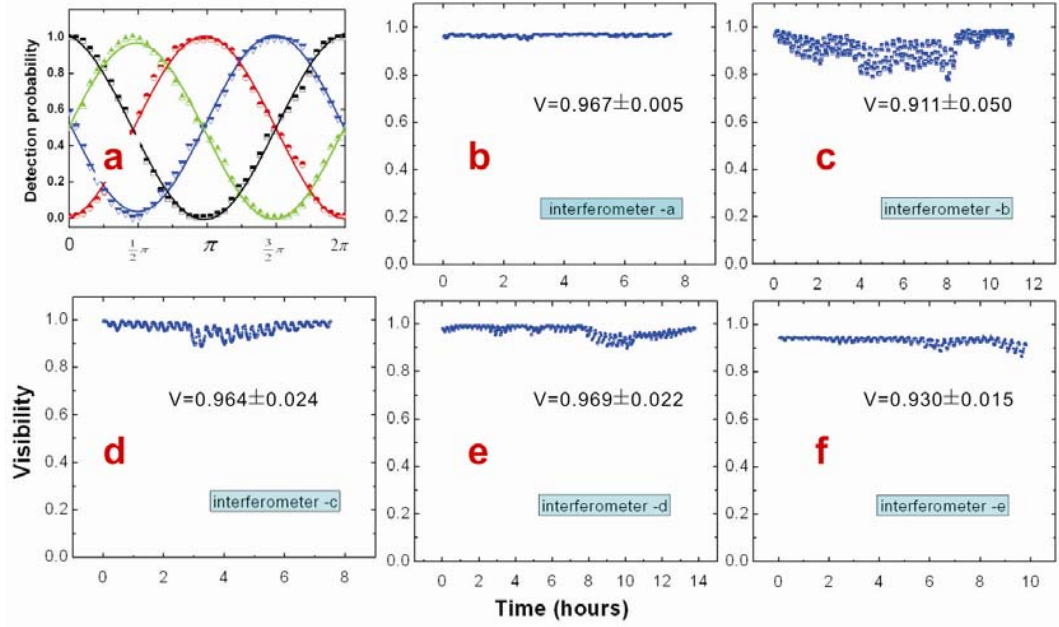

Figure A2: Test data sheet of the five Sagnac-like configuration interferometers used in the present experiment. For each interferometer, we prepare the input photon in the polarization state of $|+\rangle = (|H\rangle+|V\rangle)/\sqrt{2}$ and, after its passing through the PBS, in the EPR-like hyper-entangled state $(|H\rangle|H'\rangle+|V\rangle|V'\rangle)/\sqrt{2}$. If we project the polarization qubit into $|+\rangle$ ($|R\rangle = (|H\rangle+i|V\rangle)/\sqrt{2}$), the probability to detect the spatial qubit in the state $(|H'\rangle \pm e^{i\theta}|V'\rangle)/\sqrt{2}$ will in principle be $(1\pm\cos\theta)/2$ ($(1\mp\sin\theta)/2$). The data shown in **a** are typical detection state probabilities obtained experimentally by setting the polarization state in $|+\rangle$ (black and red), or $|R\rangle$ (green and blue), and the spatial qubit in $(|H'\rangle+e^{i\theta}|V'\rangle)/\sqrt{2}$ (black and green), or $(|H'\rangle-e^{i\theta}|V'\rangle)/\sqrt{2}$ (red and blue) with $\theta$ ramping through $2\pi$. The fitted curves agree well with the theory, exhibiting very high visibilities (defined as $V = [P(|+\rangle|+'\rangle)-P(|-\rangle|-'\rangle)]/[P(|+\rangle|+'\rangle)+P(|-\rangle|-'\rangle)]$, where $|+'\rangle = (|H'\rangle+|V'\rangle)/\sqrt{2}$). The long-term observations of the visibilities of the five interferometers (over 8-12 hours) are shown in **b-f** together with calculated average values and standard deviations, which prove the stability of these interferometers.

**Entanglement witness construction with local filter**

In order to investigate the entanglement contained in the ten-qubit state further, we make use of local filtering operations [3,4]. That is, starting from the experimentally generated state $\rho_{\exp}$ we consider the state

$$\tilde{\rho} = N \cdot F\rho_{\exp}F^+ \qquad (1)$$

where $N$ denotes a normalization and

$$F = F_1 \otimes F_2 \otimes F_3 \otimes \ldots \otimes F_{10}$$

with

$$F_i = (1+\lambda_i)|0\rangle\langle 0| + (1-\lambda_i)|1\rangle\langle 1|$$

and $-1 < \lambda_i < 1$ is a local filtering operator. Since the $F_i$ are invertible, these filter operations keep the entanglement properties, that is, they map separable (or entangled) states on separable (or entangled) ones. Therefore, by varying the parameters $\lambda_i$ one can try to find states, where fidelity of the cat state is higher or where the entanglement becomes more clearly manifest through a more negative value of an entanglement witness.

Let us first consider the optimization of the witness via this method. We consider the filtered witness $W_F = N' \cdot F^+ W F$ where $W = \frac{1}{2} - F(|Cat\rangle)$ is the witness based on the fidelity and the normalization $N'$ is such that $Tr(W_F) = Tr(W)$ to make the witnesses' mean values comparable. Note that the filters $F_i$ are chosen in such a way that the mean value of $W_F$ can be evaluated from just the same measurement data as $W$. Optimizing over the $\lambda_i$ yields a filtered witness with

$$\langle W_F \rangle = -0.072 \pm 0.018,$$

improving the violation about one standard deviation.

Next, we consider an application of the filter directly to the state as in Eq. (1). Here, we ask to which extent the fidelity of the state can be improved using local filters as above. Note that such filter operations are not only a theoretical consideration; they can be implemented using partial polarizers [3] (for the polarization qubits) or beam splitters with a variable reflection (for the spatial qubits). Also, as now the normalization of the state is fixed, the optimization is not the same as above. One obtains a filter yielding the fidelity

$$F(|Cat\rangle^{10}) = 0.581 \pm 0.020.$$

Again, the special form of the filter guarantees that the fidelity of $\tilde{\rho}$ and the normalization $N$ can be computed, although state tomography has not been done.